\documentclass{ws-ijmpa}

\begin{document}

\markboth{M.C. Bento, R. Gonzalez Felipe and N.M.C. Santos}
{Instructions for Typing Manuscripts (A simple quintessence model)}

%
\catchline{}{}{}{}{}
%

\title{A SIMPLE QUINTESSENCE MODEL}

\author{M.C. BENTO}

\address{Centro de F\'{\i}sica Te\'orica de
Part\'{\i}culas, Instituto Superior T\'{e}cnico, Avenida Rovisco
Pais\\ 1049-001 Lisboa, Portugal\\
bento@sirius.ist.utl}

\author{R. GONZ\'{A}LEZ FELIPE}

\address{Instituto Superior de Engenharia de Lisboa, Rua Conselheiro Em\'idio Navarro\\ 1959-007 Lisboa,
Portugal\\
gonzalez@cftp.ist.utl.pt}

\author{N.M.C. SANTOS}

\address{Centro de F\'{\i}sica Te\'orica de
Part\'{\i}culas, Instituto Superior T\'{e}cnico, Avenida Rovisco
Pais\\ 1049-001 Lisboa, Portugal\\
ncsantos@cftp.ist.utl.pt}

\maketitle

\begin{history}
\received{Day Month Year}
\revised{Day Month Year}
\end{history}

\begin{abstract}
A simple model of quintessential inflation with the modified exponential potential $e^{-\alpha \phi} \left[A+\left(\phi-\phi_0\right)^2\right]$ is analyzed in the braneworld context. Considering reheating via instant preheating, we conclude that the model exhibits transient acceleration at late times for $0.96 \lesssim A \alpha^2 \lesssim 1.26$ and $271 \lesssim \phi_0\, \alpha \lesssim 273$, while permanent acceleration is obtained for $2.3\times10^{-8} \lesssim A \alpha^2 \lesssim 0.98$ and $255 \lesssim \phi_0\, \alpha \lesssim 273$. The steep parameter $\alpha$ is constrained to be in the range $5.3 \lesssim \alpha \lesssim 10.8$.
\end{abstract}

\ccode{PACS numbers: 98.80.Cq, 95.36.+x, 04.50.-h}

\section{Introduction}

The Universe seems to exhibit an interesting symmetry with regard to the accelerated expansion, namely, it underwent inflation at early epochs and is believed to be accelerating at present.It is then natural to ask whether one can build a model to join these two ends without disturbing the thermal history of the Universe. Attempts have been made to unify both these concepts  in which a single scalar field plays the role of the inflaton and quintessence - the so-called quintessential inflation.

On the other hand, in recent years there has been increasing interest in the cosmological implications of a certain class of braneworld scenarios where the Friedmann equation is modified at very high energies. In particular, in the Randall-Sundrum type II (RSII) model~\cite{Randall:1999vf} the square of the Hubble parameter, $H^2$, acquires a term quadratic in the energy density,
\begin{equation}
H^2 = {1 \over 3\, M_4^2}\, \rho\, \left[1 + {\rho \over 2 \lambda}\right]~.
\label{eq:Friedmann}
\end{equation}
allowing slow-roll inflation to occur for potentials that would be too steep to support inflation in the standard Friedmann-Robertson-Walker (FRW) cosmology.
 where $M_4$ is the 4D reduced Planck mass and  $\lambda$ is the brane tension.

 In this paper we show that the modified exponential potential (hereafter we adopt natural units, $M_4=1$, unless stated otherwise)
\begin{equation}
V(\phi)=e^{-\alpha \phi} \left[A+\left(\phi-\phi_0\right)^2\right]
\label{eq:albrecht}
\end{equation}
also leads to a successful quintessential inflation model.

In the context of quintessence, this potential was first analyzed by Albrecht and Skordis (AS)~\cite{Albrecht:1999rm}.

  The model displays an interesting feature: it can lead to both permanent and transient acceleration regimes.

 These models belong to the category of nonoscillating models in which the standard reheating mechanism does not work. In this case, one can employ instant preheating. This mechanism is quite efficient and robust, and is well suited to nonoscillating models~\cite{Felder:1999pv}.

\begin{figure*}[t]
\centerline{\psfig{file=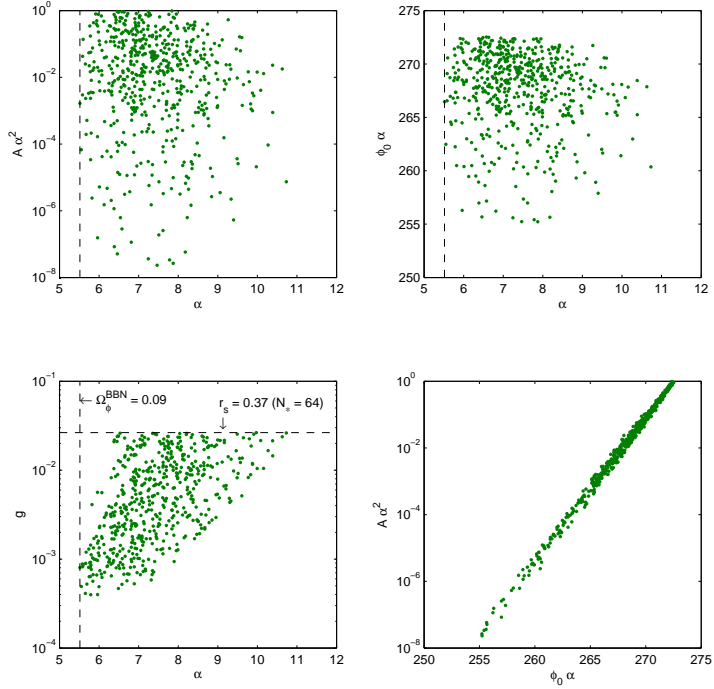,width=11cm}}
\caption{Parameter space consistent with all the observational constraints considered, for the permanent acceleration case.} \label{fig:permanent}
\end{figure*}

\begin{figure*}[t]
\centerline{\psfig{file=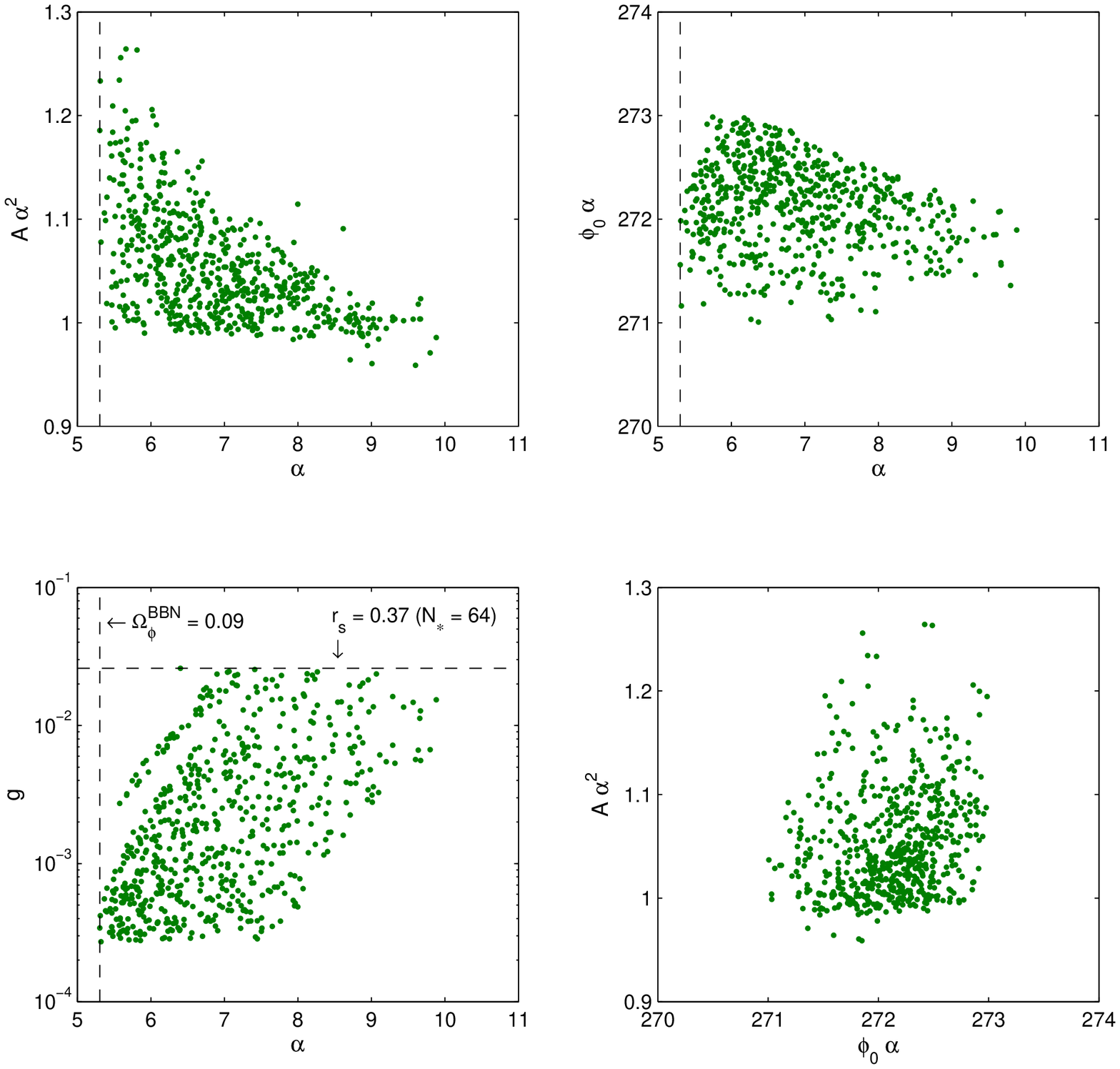,width=11cm}}
\caption{Same as in Fig. \ref{fig:permanent}, but for the transient acceleration case.} \label{fig:transient}
\end{figure*}

\section{Late Time Evolution and observational constraints}

In our study, we perform a random analysis on the potential parameters $\alpha$, $A$, and $\phi_0$, together with $N_\star$ and  $g$. We consider the two possible late time behaviors: permanent or transient acceleration.

 A stringent bound comes from the amount of dark energy during nucleosynthesis $\Omega_\phi^{\rm BBN} (z \simeq 10^{10}) \lesssim 0.09$~\cite{Bean:2001wt}. At present, we consider the following conservative bounds:
\begin{equation}
0.6 \leq  h \leq 0.8~, \quad
0.6 \leq  \Omega_\phi^0 \leq 0.8~,
w_\phi^0 \leq -0.8~,\quad
 q_0 < 0~,
\end{equation}
where $q \equiv - \ddot{a}/(a\,H^2)$ is the deceleration parameter.

The results of our analysis are displayed in Figs.~\ref{fig:permanent} and \ref{fig:transient}.
The upper bound on the coupling $g$, coming from the $r_s$ constraint on $N_\star$, as well as the lower bound on the potential parameter $\alpha$, resulting from the bound on the amount of dark energy during BBN, is also shown in the figures (horizontal and vertical dashed lines, respectively). Decreasing $g$ or increasing $\alpha$ prolongs the kinetic regime. If this regime is too long, the history of the Universe is spoiled.  This allows us to put a lower and an upper bound on $g$ and $\alpha$, respectively. From the complete numerical analysis we find that the model exhibits transient acceleration at late times for
\begin{align}\label{eq:boundtrans1}
5.3 &\lesssim \alpha \lesssim 9.9\,,~~
2.7 \times 10^{-4} \lesssim g \lesssim 2.6\times10^{-2}\,,
\end{align}
\begin{align}\label{eq:boundtrans2}
0.96 \lesssim A \alpha^2 \lesssim 1.26~,~~
271 \lesssim\, \phi_0 \alpha \lesssim 273~,
\end{align}
while permanent acceleration is obtained for
\begin{align}\label{eq:boundperm1}
5.5 \lesssim \alpha \lesssim 10.8\,,~~
4.0 \times 10^{-4} \lesssim g \lesssim 2.6\times10^{-2}\,,
\end{align}
\begin{align}\label{eq:boundperm2}
2.3\times 10^{-8} \lesssim A \alpha^2 \lesssim 0.98~,~~
255 \lesssim\, \phi_0 \alpha \lesssim 273~.
\end{align}

The number of $e$-folds from horizon crossing till the end of inflation $N_\star$ and the value for the 5D Planck mass are very constrained:
\begin{align}\label{eq:boundNstarf}
64 \lesssim N_\star \lesssim 66~\,,~~9.0 \times 10^{-4} \lesssim \dfrac{M_5}{M_4} \lesssim 1.9 \times 10^{-3} ~,
\end{align}
which imposes strong constraints on the inflationary observables $n_s$ and $r_s\,$.

\section{Conclusions}

We have analyzed a simple model of quintessential inflation in the RSII braneworld context with a  modified exponential potential. Assuming that the Universe was reheated via the instant preheating mechanism, we have shown that the evolution of the scalar field from inflation till the present epoch is consistent with the observations in a wide region of the parameter space. Requiring that the model meets various cosmological constraints at the different stages of the evolution, we were able to constrain tightly its parameters, as summarized in Eqs.~(\ref{eq:boundtrans1})-(\ref{eq:boundNstarf}).

\section*{Acknowledgments}
N.M.C.S. acknowledges the support of the Funda\c{c}\~{a}o para a Ci\^{e}ncia e a Tecnologia (FCT, Portugal) under the Grant No. SFRH/BPD/36303/2007. This work was also partially supported by FCT through the Project No. POCTI/FIS/56093/2004.


\end{document}